\documentstyle[12pt]{article}
\begin{document}
% definitions

\newcommand{\newc}{\newcommand}

\newc{\be}{\begin{equation}}
\newc{\ee}{\end{equation}}
\newc{\ba}{\begin{eqnarray}}
\newc{\ea}{\end{eqnarray}}
\newc{\bea}{\begin{eqnarray}}
\newc{\eea}{\end{eqnarray}}
\newc{\D}{\partial}
\newc{\ie}{{\it i.e.} }
\newc{\eg}{{\it e.g.} }
\newc{\etc}{{\it etc.} }
\newc{\etal}{{\it et al.} }
\newc{\ra}{\rightarrow}
\newc{\lra}{\leftrightarrow}
\newc{\no}{Nielsen-Olesen }
\newc{\lsim}{\buildrel{<}\over{\sim}}
\newc{\gsim}{\buildrel{>}\over{\sim}}
\begin{titlepage}
\vspace*{-2.5cm}
\hspace*{9cm}October 1998\\
%\hfill  \\ \hfill  
\hspace*{9.5cm} hep-th/9810230\\
\hspace*{9.5cm} DEMO-98/03\\
\hspace*{9.5cm} CERN-TH/98-279\\
\vskip 1cm
\begin{center}

{\large {\bf NON-ABELIAN  Q-BALLS IN SUPERSYMMETRIC THEORIES}}

\vskip .4in
{\large Minos Axenides},\footnote{
axenides@gr3801.nrcps.ariadne-t.gr }  {\large Emmanuel Floratos}\footnote{
manolis@cyclades.ariadne-t.gr}\\[.1in]
 {\em Institute of Nuclear Physics,\\ N.C.S.R.  Demokritos \\
153 10, Athens, Greece}\\
\vskip .15in
 and\\
\vskip .15in
{\large Alexandros Kehagias}\footnote{
Alexandros.Kehagias@cern.ch},\\[.1in]
{\em Theory Division, CERN, CH-1211 Geneva 23, Switzerland}\\

%\newpage
\vskip .4in
%\begin{ABSTRACT}
{\bf ABSTRACT}
\end{center}

\vspace{0.8cm}

\noindent
We demonstrate the existence of non-abelian non-topological 
solitons such as Q-balls in the spectrum of
Wess-Zumino models with non-abelian global symmetries. 
We conveniently name them Q-superballs and 
identify them 
for short as  Q-sballs. 
More specifically, we show that in contrast to the
non-supersymmetric case, they arise in renormalizable 
potentials with cubic self-interactions 
of only one dimensionful parameter and
 for the entire parameter space of the model available. 
We solve the field equations  and present
the explicit form of the Q-sball solution. We compute 
its main physical properties and
observe that  in the supersymmetrically invariant vacuum
Q-sballs form domains  of manifestly broken sypersymmetry.
%\end{abstract}
\end{titlepage}

\section{\bf INTRODUCTION AND CONCLUSIONS}
Non-topological solitons are extended finite energy configurations that 
arise in $3+1$ dimensional field
theories that contain scalar fields with unbroken global symmetries
\cite{R,FLS,LP}. In the presence of attractive
scalar self-interactions in their potential energy and a conserved global
 charge they admit non-dissipative
time dependent solutions to their equation of motion with their total 
binding energy per unit charge
$E/Q < \mu$ being smaller than the perturbative mass scale of free mesons. 
The total energy of such
configurations scales with the conserved charge Q as follows:
\bea
E \propto Q^s \, ,~~ s\leq 1\, . 
\eea
They are spherically symmetric configurations of scalar fields which 
monotonically decrease to zero at
spatial infinity.
They receive both surface and volume contributions to their total energy 
for an arbitrary
value of Q with the latter dominating for sufficiently large values of Q.
The resulting soliton is a homogeneous blob of matter possessing a very 
thin skin which can be neglected for
all practical purposes. It has been called a Q ball \cite{C}. More 
importantly it exists in the strict 
thermodynamic limit $ V\ra \infty, E/Q=const$.
Such scalar field coherent configurations could also 
carry additional 
charges giving rise to long range gauge fields \cite{FLS,A,KST}.

In the low energy world strong interactions alone respect strangeness 
and isospin. In the context
of effective Lagrangians the possible existence of short lived Strange 
and Isospin balls as resonances in the
low energy QCD spectrum has been discussed for some time now \cite{DHS,A}.

In the minimal electroweak theory the pretty well respected baryon and 
lepton quantum numbers don't
allow for  such configurations. Recently, however, in supersymmetric 
extensions of the standard model
such as the MSSM \cite{K,KS} the existence of baryon and lepton balls 
was demonstrated.
They are composed of charged squarks and sleptons along with Higgs scalar 
degrees of freedom. 
In what follows
we will conveniently call a Q-ball which contains a supersymmetric scalar 
matter component such   such
as sfermions a Q-superball denoting it for short Q-sball.
Abelian Q-sballs typically arise through soft susy breaking interactions
\cite{K} such as 
non-renormalizable polynomial scalar self interactions that
appear in the flat directions of the supersymmetric theories \cite{EM,DRT}. 
They can also be generated from
SUGRA induced quantum logarithmic corrections \cite{EM}.
It has been known from earlier work that the distinguishing characteristic 
of non-abelian Q-balls, such
as $SO(3)$ or $SU(3)$, is that they can arise through solely renormalizable 
potentials that contain
cubic interactions \cite{FLS,A,SCA}. 
More precisely we consider a scalar model 
given by  
\be
{\cal L} \,\,=\,\,\frac{1}{2} Tr (\partial_{\mu}\phi)^2\,\,-\,\,Tr U(\phi)\, , 
\ee
and
\be
U(\phi)\, =\, \frac{1}{2} \mu^2\phi^2\, +\, \frac{g\phi^3}{3!}\,+\, 
\frac{\lambda\phi^4}{4!}\, , 
\ee
with $\mu^2 >0$ and $\lambda>0$. 
Here the field $\phi$ transforms according to the adjoint representation of 
$SU(3)$ $\phi \rightarrow u \phi u^{\dagger}$. Identically it
can be taken to consist of five real valued scalar fields transforming 
according to the spinor 
representation of $SO(3)$ with the above potential being the most general 
$SO(3)$-invariant, 
renormalizable interaction.\footnote{As our  arguments  do not get modified 
in the presence of additional
gauge fields and symmetries we restrict ourselves to scalar theories.}

The conserved charges of both models are assembled into a traceless
hermitian matrix 
\be
Q\,=\,i\,\int\,\,d^3 x \left[ \dot{\phi},\phi\right]\, .
\ee
Both non-abelian models possess 
an identical set of defining equations provided 
that u and $\phi$ are pure real and thus Q is a pure
imaginary. For a renormalizable potential with two independent dimensionful 
couplings  $\mu$ and  $g$ 
along with one dimensionless $\lambda$  Q-balls were shown to be present for 
a window in the space of
free parameters given by 
\be
1 \leq \beta\equiv \frac{g^2}{\mu^2}{\lambda} \leq 9\, .
\ee
The upper bound guarantees that the state $\phi=0$ is the unique vacuum of 
the model while the lower
bound secures that the Q-balls does not decay into free mesons.

In the present work we explore the effect supersymmetry might have in the 
above picture
by investigating the spectrum of global supersymmetric models that give rise 
to potentials
of the same form as above. We establish the
presence of Q-sballs in the spectrum of Wess-Zumino \cite{WB} supersymmetric 
models 
of a single chiral superfield
and non-abelian global symmetries. 
We find that supersymmetry constrains the freedom between couplings in such a 
way that the resulting
potential is characterized by a single dimensionful coupling $m$ and a 
dimensionless one $h$ 
and it possesses 
\be
\beta=3 \, .
\ee
In effect Q-sballs are present in its spectrum for every value of its free
 parameters.
We solve analytically the equations of motion exhibiting explicitly a Q-sball
solution and compute
its energy and charge densities. We observe that in a supersymmetrically 
invariant vacuum the
Q-sball interior breaks supersymmetry explicitly. The supersymmetry breaking 
scale is set
by the internal rotational frequency of the superballs. Certainly in the 
presence of additional
soft susy breaking terms in the potential the above observation does not 
hold as the vacuum breaks
 supersymmetry as well. It is in this very respect that our present study 
makes apparent the distinctive
characteristics of non-abelian Q-sballs that arise through solely 
renormalizable interactions.

Our Q-sball solution is a minimal one as it has  $det Q=0$. As such it is an 
equally good solution
for both $SO(3)$ and $SU(3)$ models. Non-minimal Q-sball configurations 
with  $det Q\neq 0$ are expected to exist for the
$SU(3)$ model \cite{SCA,FLS}. This is because the Q matrix can be unitarily 
rotated into a diagonal
form ${\it{diag} [q_1,q_2 -(q_1 +q_2)]}$ with $q_1\cdot q_2 >0$. 
Consistently with Q conservation
\be
diag (q_1,q_2,-(q_1+q_2))\,=\,diag(q_1,0,-q_1)\,+\,(0,q_2,-q_2)\, ,
\ee
it has been shown that it is energetically favourable for a large Q-ball with 
$det Q\neq 0$ to
fission into two minimal ones with $det Q=0$. The existence of a minimal 
Q-sball automatically implies
the presence of the same phenomenon of fission in the $SU(3)$ model as well.
Our conclusions, as they apply to the scalar sector of a more 
general theory,  go through in the presence of additional  
non-renoramalizable  interactions as well as 
gauge fields and charges that 
our Q-sball configuration might carry. 
Our present study suggests the presence of non-topological solitons 
in the spectrum 
of any supersymmetric gauge field theory with residual non-abelian 
global symmetry and appropriate
attractive interactions.
  
\section{\bf MINIMAL  AND MAXIMAL  Q-SBALLS IN $SU(3)$}
We consider the Wess-Zumino Lagrangian \cite{WB}
\bea 
{\cal L} &=& i \D_{m}\bar{\psi}_i \bar{\sigma}^{m}\psi_i - 
 \nabla_{m}\phi^{i*}\nabla^{m}
\phi^i - \frac{1}{2} \frac{\partial^2W}{\partial\phi^i\partial\phi^j}
\psi_i \psi_j\nonumber \\&& - \frac{1}{2} 
\frac{\partial^2W^{*}}{\partial\phi^{i*}
\partial \phi^{j*}}
\psi_i \psi_j - 
U(\phi,\phi^{*})\, ,  
\eea
where $\phi$ and $\psi$ are the dynamical scalar and fermion components of
an  $N=1$ chiral multiplet and $W$ is the superpotential. The potential  
is given in terms of the superpotential as
\be
U=\sum_i\left|\frac{\D W}{\D \phi^i}\right|^2\, .
 \ee
  %\be
  %\Xi = \phi + \sqrt{2}\theta\psi + \theta\theta F
  %\ee
  %Here F is an auxiliary field which can be eliminated through the
  %Euler equations.
  The above model obeys the following supersymmetry transformations
\bea
\delta_{\xi}\phi &= &\sqrt{2}\xi \psi \, ,
\nonumber\\ \delta_\xi\psi &=& i\sqrt{2}\sigma^m
\bar{\xi}\D_m\phi \, .
%= \sqrt{2}\xi F\nonumber 
%\\ \delta_\xi F&=& i\sqrt{2}\bar{\xi}
%\bar{\sigma}^m \D_m\psi
\eea
As the Lagrangian is explicitly supersymmetric it forces the fermions
and scalar components to be degenerate in mass. 
In what follows we will focus in 
the scalar sector of the model. We will assume that the scalars belong to 
the adjoint representation of a global symmetry group $G$. For illustrative 
purposes, we will take $G$ to be $SU(N)$ in which case, $\phi$ is an 
$N\times N$ hermitian and traceless matrix. The most  general superpotential
which leads to renormalizable interactions is given by
\bea
{\cal{ W}}\,&=&\,m Tr \phi^2\,+\,h Tr \phi^3 \nonumber \\
            &=&\,m \phi_{ij}\,+\,h\,\phi_{ij}\phi_{jk}\phi_{kl}\, . 
\eea
As a result, the potential of the model  model 
takes the form
\be
U\,=\, 4m^2 Tr \phi^2\,+\, 12m h Tr \phi^3 \,+\, 9 h^2 Tr \phi^4\, .
\ee

We observe that in the general parametrization of the introduction,  
$\mu^2 = 8 m^2$, $g = 72 m h$ and $ \lambda = 216 h^2 $ 
\be
  \beta\,=\, 3\, .
\ee
Remarkably the supersymmetric structure of the model while preserving the 
available number
of renormalizable self interactions intact, it reduces the number of 
independent dimensionful
parameters from two $\mu$ and $g$ in the non-susy case to one $m.$ in the 
present supersymmetric
one. In effect it is associated to a theory which
possesses a single value for the dimensionless ratio  $1<\beta=3<9$ for
 which
previous analysis suggests that it admits Q-balls in its spectrum for all 
values of its parameters $m$ and $h$.

In what follows we will utilize the available formalism \cite{SCA} to
obtain the precise Q-sball configuration.
We will focus in the $SU(3)$ model although our formalism applies to the 
$SU(N)$ case as well.
We are interested in initial value data $\phi$ and $\Pi\equiv \dot{\phi}$ at
 a fixed time, 
which minimize the
energy keeping Q fixed. Such data obey
\be
\delta \left [ H + Tr \omega Q \right ]\,=\, 0\, ,
\ee 
where H is the energy written as a functional of $\phi$ and $\Pi$ at fixed 
time, 
and $\Omega$ is a Lagrange multiplier, a traceless hermitian $ 3\!\times\!3$
matrix. 
The above equation implies that
\be
\frac{\delta H}{\delta \phi} \,=\, -i \left [ \Omega, \Pi\right]\, , 
\ee
and
\be
\frac{\delta H}{\delta \Pi}\,=\, i\,\left[ \Omega, \phi\right]\, .
\ee
We can observe that $\phi$ and $\Pi$ are initial value data for a solution of 
the equation of motion that
is in steady rotation in internal space
\be
\phi (x,t)\,=\, e^{ {i\Omega t}} \phi(x,t) e^{ {-i \Omega t}}.
\ee
By restricting ourselves to large Q-balls, we consider a sphere of volume V in 
the interior of which
$\Pi$ and $\phi$ are constant. In other words we neglect the contribution to 
$Q$  and the energy from the
surface region where $\Pi$ and $\phi$ go to zero.
The energy is then given by
\be
E\,=\, V Tr\,\left(\,\frac{1}{2}\Pi^2 \,+\,U \right)\, , 
\ee
while
\be
Q\,=\,iV [\Pi,\phi]\, .
\ee
In order to firstly look for minimal Q-balls with $det Q=0$ , i.e. 
those for which one eigenvalue of Q
vanishes it is convenient to make a unitary transformation so 
that $\phi$ is diagonal 
\be
\phi\,=\, diag (\phi_1,\phi_2,\phi_3),
\ee
with the eigenvalues taking the following order
\be
\phi_1 \geq \phi_2 \geq \phi_3.
\ee
When $\phi$ is diagonal, $\Pi = 
i[\omega,\phi]$ and $ Q=i[\phi,\dot{\phi}]$ take a pure off-diagonal form
\be 
\Pi_{ij}=i\Omega_{ij}(\phi_j - \phi_i)\, ,  
\ee
and
\be
Q_{ij}= - V\Omega_{ij}(\phi_j - \phi_i)^2\, .
\ee
Thus we can write E in terms of Q and $\phi$ as
\be
E\,=\, \sum_{i>j} \frac{ | Q_{ij}|^2}{V(\phi_i -\phi_j)^2}\,+\, V Tr U\, .
\ee
By varying V in the energy expression by keeping $\phi$ and Q fixed we find
\be 
Tr \frac{1}{2} \Pi^2 \,=\,Tr U\, .
\ee
Using the equations of motion
\be 
\ddot{\phi}=[\omega,[\Omega,\phi]]\, ,
\ee
it can be easily shown that 
\be
Tr( \phi \frac{\D U}{\D\phi})\,=\, Tr (\ddot{\phi}\phi)\,=\,Tr \Pi^2\, .
\ee
Hence by eliminating $Tr \Pi^2$ from eqs(25,27) and  we get
\be 
Tr (\phi \frac{\D U}{\D\phi})\,=\,2 TrU\, .
\ee
For the general renormalizable potential given before, 
the above relation takes the form
\be 
Tr( 2g\phi^3 + \lambda\phi^4)\,=\,0\, .
\ee
We now proceed to determine the Q-sball configuration in detail by the 
method utilized for a general
potential\cite{SCA}.

We may choose to minimize the energy by taking $Q_{13}=Q_{31}^*=iq$ a 
positive imaginary and the only
nonzero matrix element of Q. Since it is a constant of motion it must 
commute with $\omega$.
With no loss of generality we choose 
the $\Omega$ matrix in complete similarity with the Q matrix
to have only one nonzero off diagonal element 
\be
\Omega_{13}=\Omega_{31}^*= -i\omega\, .
\ee
Consequently,
\be
\omega = \frac{q}{V (\phi_1 -\phi_3)^2}.
\ee
We see that $\phi$ is real at all times. In other words every minimal 
$SU(3)$ Q-ball is unitarily
equivalent to an $SO(3)$ one as we asserted before.
Rewriting the equations of motion as before for 
a renormalizable potential U and the chosen $\omega$ and $\phi$
we get
\begin{eqnarray}
\mu^2\phi + \frac{1}{2} g\phi^2 + 
\frac{1}{6}\lambda\phi^3 - \frac{1}{3}Tr(\frac{1}{2}g\phi^2+\frac{1}{6}
\lambda\phi^3) I\,=\,[\Omega,[\Omega,\phi]]\nonumber \\
= 2\omega^2(\phi_1-\phi_3) \left(\begin{array}{rrr}1&0&0\\0&0&0\\
0&0&-1\end{array}\right)\, .
\end{eqnarray}
We note that the ($22$) element of 
the above equation contains no reference to $\omega$. Along with eq.($29$)
we can solve them to obtain the two 
independent eigenvalues of $\phi$. Once we have these two, we
can use them in the remaining of eq.($32$) 
to determine $\omega$. Finally once we have 
$\omega$ we can determine V from Q
in eq.$(31)$.
To this end we adopt the parametrization of 
\cite{SCA} for the eigenvalues of $\phi$ in terms of
$\phi_2$ and a dimensionless variable y as follows:
\be
\phi_1= -\frac{1}{2}\phi_2(1+y), \,\,\,\phi_3=-\frac{1}{2}\phi_2(1-y)\, .
\ee
By changing the sign of $y$ the values of $\phi_1$ and $\phi_3$ get 
exchanged with no 
effect in our equations.
With no loss of generality we can thus take $y$ to be positive.
In terms of these variables eq.($29$) takes the form
\be
\frac{3g}{2}\phi_2^3(1-y^2) + \frac{1}{8}\lambda\phi_2^4(3+y^2)^2=0\, .
\ee
The ($22$) element of eq.($32$) is identified to be
\be
\mu^2\phi_2 + \frac{1}{12}g\phi_2^2 (3-y^2) + 
\frac{1}{24}\lambda\phi_2^3(3+y^2)=0\, .
\ee
Substituting into it the expression for $\phi_2$
\be
\phi_2=\frac{12g(y^2-1)}{\lambda(3+y^2)^2}\, ,
\ee
we obtain a cubic equation for $z=y^2$ which if expressed in terms 
of the parameters $m,h$ takes the form
\be
z^3 - 15z^2 -9z -9 =0\, .
\ee
There exist two imaginary roots and one real 
one which is relevant for us and it is 
\be
z\approx 15.613\, .
\ee
In terms of $z$, $\phi_2$ is given by 
\be
\phi_2=4\frac{m}{h}\frac{z-1}{(3+z)^2}\approx 0.169\frac{m}{h}\, .
\ee
It automatically fixes the other two components $\phi_1$ and $\phi_2$ 
through eq.(33) giving
\be
 \phi_1 = -0.418 \frac{m}{h} \,\,\, ,\,\,  \phi_3  = 0.249 \frac{m}{h}\, .
\ee

Thus a minimal Q-sball configuration  with $det Q =0$ is 
fully determined in terms of $m$ and $h$.
We can now proceed to specify its physical properties.
The rotational frequency of the Q-sball is determined to be
\be
\omega^2=\frac{g}{48} \phi_2 (z-9)=6\frac{(z-1)(z-9)}{(3+z)^2}m^2\approx 1.67 m^2\, ,
\ee

We observe that $\omega^2 < \mu^2 =8 m^2 $, hence the 
binding  meson energy is smaller than the corresponding free meson mass
which is the statement of existence of a coherent scalar bound state in the
spectrum of the model.  
The charge density is given by 
\be
\frac{q}{V} =  \omega \phi^{2}_{2} \sqrt{z} =
154.85  \frac{(z-1)^{5/2}(z-9)^{1/2}}{(3+z)^5}\frac{m^3}{h^2}\, .
\approx 0.145 \frac{m^3}{h^2}
\ee

Non-abelian minimal Q-sballs are thus generically present in the 
scalar spectrum of a Wess-Zumino model with a non-abelian global symmetry.
We argue that in the presence of degenerate in mass
fermionic fields $\psi$ which are components of the chiral superfield  
our scalar Q-ball is stable. Indeed the global charge is
not favourable to be radiated away from the surface of our configuration
by the fermions as long as supersymmetry in the vacuum is manifest.
The {\it evaporation } of the Q-sball \cite{CCGM} will only take
place in the presence of fermions with mass smaller than the characteristic
energy per unit charge of our soliton given by  
\be
 \frac{E}{q} = \omega = 1.292 m\, .
\ee

This can conceivably occur in the above model through the introduction
of additional non-renormalizable soft supersymmetry breaking terms
bringing down the fermion mass gap which is given by $\mu \approx 2.83 m$. 

An interesting property of the Q-sball solution is that
it breaks explicitly supersymmetry. Indeed the solution given above
violates supersymmetry since
\be
\delta_{\xi}\psi \propto 
\sigma^0\bar{\xi}\dot{\phi} \propto \sigma^0\bar{\xi} [\Omega, \phi] \neq 0\, .
\ee
In the supersymmetrically invariant theory Q-sballs are domains
where supersymmetry is explicitly broken. Fermions and bosons traversing
such configurations are not degenerate in mass in their interior.
Our discussion and demonstration of 
the existence of minimal non-abelian Q-sballs 
in Wess-Zumino type of models do not get modified in the presence also
of gauge fields \cite{A,KST} 
and additional gauge symmetries in the Lagrangian. 

We complete our demonstration of existence of 
non-abelian Q-sballs by discussing 
non-minimal  Q-sballs in the model considered. 
They arise through the minimization of the energy 
functional by keeping $det Q\neq 0$ fixed. 
We adapt the arguments of previous work 
\cite{SCA} in a straightforward manner.
For convenience we will focus in the $SU(3)$ model identifying the
 corresponding non minimal
non-dissipative solutions as maximal.
To that end we take the meson charge to be diagonal of the form
\be
Q = \left(\begin{array}{rrr}N&0&0 \\0& n-N& 0\\ 0&0&-n\end{array}\right).
\ee
The minimum energy of such a meson will be given by
\be
E \sim \mu (n + N)\, .
\ee
However as it was argued before \cite{SCA} the configuration can attain a 
lower energy
by fissioning into two minimal  Q-sball solutions. The charge decomposition
 consistently
with global charge conservation is the following
\be
Q = \left(\begin{array}{rrr}N-n & 0 & 0\\0 &n-N & 0\\ 0& 0& 0\end{array}
\right) +
    \left(\begin{array}{rrr}n &0&0 \\0 &0 & 0\\ 0& 0& -n\end{array}\right)
\ee
The energy of the configuration of two widely separated minimal Q-sballs with
the above charge assignments is reduced to
\be
 E \sim \mu N\, .
\ee
By adapting the argument to our case we expect that if Q-sballs with 
$detQ \neq 0$
exist they are energetically favorable to fission into two minimal Q-sballs
with $det Q = 0$. 

By introducing the variable $y = n/N$ we can denote departure from Q charge 
minimality.
One can numerically demonstrate that for $ 0 <y < 0.06$, i.e. for not large 
values
of $det Q$, and in a renormalizable meson theory such as is our supersymmetric
model at hand, maximal  Q-sball 
configurations exist which are favorable to fission into two minimal ones 
with charge 
of the form given above . For more details of the algebra involved one can 
consult 
previous work done \cite{SCA} on the issue.
  
\section{Acknowledgements}
One of us A.K. thanks Dr Kusenko for discussing his work with him.

\end{document}